\documentclass[preprint,nofootinbib,superscriptaddress,amsmath,amssymb]{revtex4}
\usepackage[T1]{fontenc}
\usepackage[utf8]{inputenc}
\usepackage{amsmath} 
\usepackage{graphicx} 
\usepackage{bbm}
\usepackage{etoolbox}
\preto\subequations{\ifhmode\unskip\fi}
\usepackage{comment}
\usepackage{color}
\usepackage{enumerate}
\usepackage{subfigure}
\usepackage{setspace}
\usepackage{multirow}
\usepackage{graphicx}
\usepackage{float}
\usepackage{array}

\newcommand{\be}{\begin{equation}}
	\newcommand{\ee}{\end{equation}}
\newcommand{\bi}{\begin{itemize}}
	\newcommand{\ei}{\end{itemize}}
\newcommand{\nn}{\nonumber}

\begin{document}

\title{Hidden Regimes during Preheating}
\author{Gizem \c{S}eng\"or}
\email{gizemsengor@gmail.com}
\affiliation{Department of Physics, Syracuse University, Syracuse, NY 13244, USA}
\affiliation{CEICO, Fyzik\'{a}ln\'{ı} \'{u}stav Akademie v\v{e}d \v{C}R, Na Slovance 2, 182 21 Praha 8, Czechia}
\date{\today}

\begin{abstract}
The Effective Field Theory (EFT) of Preheating with scalar fields, implies three types of derivative couplings between the inflaton and the reheating field. Two of these couplings lead to scales below which only one of the two species appear as the low energy modes. In this paper, the variety of low energy regimes in terms of the species they accommodate are explored by studying the scales introduced by the derivative couplings and the dispersion relations they lead to. It is noted that the EFT of two scalar fields can give rise to non-trivial sound speed for both the inflation and reheating sector even at scales where modes of both species propagate freely, suggesting the presence of additional heavy fields. The regimes where one of the species affects the dispersion relation of the other while not appearing as an effective mode itself, are named as ``Hidden Regimes'' during preheating. 
\end{abstract}
\maketitle
\thispagestyle{empty}
\tableofcontents
\newpage

\section{Introduction}

In single field inflation, only one type of field, the inflaton, dominates the overall energy momentum density of the universe. At the background level, this field has some time dependence which leads to a time dependent cosmological background $H(t)$. This background will not remain invariant under time diffeomorphisms. However the time diffeomorphisms are a symmetry of the action. On such a time dependent background, there exists  modes of a physical scalar perturbation which transforms nonlinearly under further time diffeomorphisms, making sure that the action itself remains invariant. These scalar modes are the inflationary scalar perturbations, observed as temperature fluctuations in the Cosmic Microwave Background (CMB). They lead to formation of protogalaxies in the early universe, on an otherwise homogeneous background. The present day universe is filled with many other types of matter, such as the particles of standard model, dark matter and dark energy. These correspond to perturbations of different types of matter fields, not all of which are scalar. It is an open question to understand how and when these fields start playing an important role in the universe overtime. It could be that a number of fields, nontrivial kinetic terms or non-minimal couplings are responsible for the underlying mechanism of inflation itself. While these cases are being more and more constrained by observations within a window of time during inflation \cite{Akrami:2018odb}, trying to understand couplings of the inflaton to other species is what is at hand for understanding the passage to other matter sources that are present in today's universe. It is likely that perturbations of noninflationary species begin to populate the universe after the end of Inflation rather then during \cite{Armendariz-Picon:2017llj}.

The period of inflation corresponds to an accelerated expansion, with a weakly time dependent background $H_I(t)$ that resembles the approximate de Sitter spacetime. Single field inflation ends when the Hubble parameter drops below the mass of the inflaton field $m_\phi$, $H_I(t_{end})\leq m_\phi$. At times of order $m_\phi t\gg 1$, the Firedmann equations give a matter dominated solution with oscillatory corrections whose frequency is set by the mass $m_\phi$. This era is referred to as preheating, during which the time dependence of the background $H_p(t)$ is strong. This time dependence works into the couplings of the inflaton field with perturbations of other species, and leads to their resonant growth leading to a more efficient production of the reheating modes compared to perturbative decay alone \cite{Kofman:1997yn}. At the end of this intermediate stage, the inflaton will decay into fields lighter then itself, giving rise to a radiation dominated phase. 

In \cite{Ozsoy:2015rna, Giblin:2017qjp}, general interactions for the perturbations of the inflaton and a scalar reheating field were studied during preheating with effective field theory (EFT) methods. It was noticed that studying these perturbations during preheating without addressing the dynamics that can give rise to the background $H_p(t)$ in the fashion of the EFT of quasi single field inflation \cite{Noumi:2012vr}, gives more insight then an EFT for the dynamics of the two species, which have been rather useful in the case of inflation \cite{Weinberg:2008hq,Assassi:2013gxa}. In this EFT approach, being associated with the species that drives the time dependence of the background, scalar inflationary perturbations $\delta g^{00}$ are Goldstone modes that nonlinearly realize time diffeomorphism invariance while reheating sector is introduced as an unspecified scalar perturbation $\chi$. This difference in their nature, leads to different interactions for the different species. How the background behavior $H_p(t)$ enters the quadratic self couplings of each species determines the resonant particle production, and the hierarchy $H_p(t)\ll m_\phi$ between the background scales, leads to hierarchies among the scales important for the dynamics of the perturbations \cite{Giblin:2017qjp}. 

In addition to interactions that lead to particle production, the EFT of preheating involves three different types of derivative couplings among the inflationary and reheating perturbations. The focus of this work are these derivative interactions. In the first work to address derivative couplings of the inflaton and a scalar reheating field, through an analysis of instability bands it was observed that the derivative coupling of interest does not promise very efficient production of reheating modes \cite{ArmendarizPicon:2007iv}. Among the three derivative EFT couplings, two of them describe energy scales. On a complimentary line to the conventional analyses of  instability bands, the main effort of the present work is to understand which of the two species of perturbations occur as the effective degree of freedom at what scales, by considering non relativistic limits and using the methods of \cite{Baumann:2011su}. The conclusion is that, with interactions of the type considered in section \ref{sec:beta1}, the effective lowest energy modes are the inflaton modes. For these type of couplings, the reheating perturbations work to adjust the dispersion relation of the inflaton rather then being present themselves. This suggests that the role of the reheating sector here is to assist the inflationary sector rather then being likely to be produced. The low energy regime of these couplings are named as the regime of Hidden Preheating, in the sense that the presence of the reheating field is hidden.  The original example of derivative couplings falls into this category. The situation is reversed with the reheating perturbations being the low energy modes whose dispersion relation is adjusted by the inflaton modes in the presence of couplings considered in section \ref{sec:beta3}. However the couplings of sections \ref{sec:beta2} and \ref{sec:beta3} give rise to further modifications in the dispersion relation that exist even at scales where the inflaton and reheating modes propagate freely. Similar to the polynomial couplings of light fields to heavy ones giving rise to corrections to the mass of the light fields at low energies, derivative couplings give rise to corrections for the dispersion relation of light modes at low scales. This suggests that these later kind of couplings signal the presence of more fields that are actually present and interact with the inflaton and the reheating sectors, but who are themselves too heavy to appear as propagating degrees of freedom.

The text is organized as follows. The Effective Field Theory formalism of cosmological perturbations and how it captures preheating is reviewed in section \ref{sec:EFTreview}. Section \ref{sec:background scales} together with appendix \ref{ap:background} review the general properties of the preheating background $H_p(t)$ and scales associated with important processes in the EFT. The main consequences of the derivative couplings are explored in section \ref{sec:Hidden} and section \ref{sec:conc} summarizes the main results. 

\section{Review of EFT of Cosmological Perturbations}
\label{sec:EFTreview}

Consider cosmological backgrounds as determined by the behavior of the Hubble parameter, $H(t)$, at each epoch. From inflation to dark energy domination, each one of the cosmological backgrounds posses time dependence of different strength. From the pursuit of particle physics, some field content that contributes as the energy momentum source will be responsible for this time dependence. The inflaton, $\phi_0(t)$ for example, is one such field that contributes as a homogeneous scalar source during and at the end of inflation. The Hubble parameter is related to the field content through the Friedmann equations, which in the case of a single scalar field read
\begin{subequations}
	\begin{align} 6m^2_{pl}H^2&=\dot{\phi}_0^2+2V(\phi_0)\\
	2m^2_{pl}\dot{H}&=-\dot{\phi}_0^2.
	\end{align}
\end{subequations}
Due to their time dependence, such backgrounds do not respect time diffeomorphism invariance. There exists a scalar degree of freedom $\delta\phi(\vec{x},t)$ or $\delta g^{00}$, associated with such a background, that keeps track of the difference in the rate of expansion between two points in space depending on how the time coordinate is defined in each point \cite{Hawking:1982cz}. This scalar degree of freedom transforms nonlinearly under time diffeomorphisms. As $t\to \tilde{t}=t+\xi^0$, the scalar perturbations transform according to
\begin{subequations}\label{timediffofscalar}\begin{align}
	\Delta\delta\phi&=-\xi^0\dot{\phi}_0,\\
	\Delta\delta g_{00}&=-\left[2\bar{g}_{00}\dot{\xi}^0+\dot{\bar{g}}_{00}\xi^0\right].
	\end{align}
\end{subequations}
Equations \eqref{timediffofscalar} imply that the time coordinate can be chosen such that the scalar perturbation is adjusted to appear completely in the metric via $g^{00}$ or as a field perturbation $\delta\phi$, but it can never be completely set to zero. This degree of freedom is the Goldstone boson that transforms under time diffeomorphisms in such a way so that the action on the whole remains invariant, that is the parameter $\xi^0$ never appears after a transformation.

Each epoch posses such a Goldstone mode associated to the type of field that dominates the energy momentum density. The first effective field theory to generalize the interactions of such a Goldstone mode at low energies was developed to study violations of the null energy condition \cite{Creminelli:2006xe}.  Understanding energy transfer from the matter source that dominates the background energy momentum density to another matter species, entails two different scalar fields. In the early stages of preheating, the inflaton is the dominant field and hence the Goldstone mode encodes the perturbation to the inflaton field. The other matter source, which is considered to be another scalar field $\chi$ here, can be introduced at the level of perturbations alone. This difference works into the interactions that can be written down for the preheating perturbation and the Goldstone mode. $\chi$ enters the action starting from second order in perturbations, and only inflationary perturbations contribute to $g^{00}$ at linear order.

The gauge in which the inflaton perturbation appears as part of the metric is referred to as the unitary gauge or the $\zeta$-gauge. Once this gauge is fixed the remaining symmetries are spatial diffeomorphisms. Because within this gauge, time direction is fixed and will not be changed further, the time related index zero can appear explicitly, for instance via $g^{00}$. Such terms with zero indices will transform as scalars under further spatial diffeomorphisms. By making this choice the temporal and spatial indices are separated. Thus spatial indices can also appear without accompanying temporal indices, provided the terms they appear in respect the remaining invariance under spatial diffeomorphims. This means perturbations to extrinsic curvature $\delta K_{ij}$, which transforms as a tensor under spatial diffeomorphisms, can contribute starting from second order via its trace $\delta {K^i}_i$. With these concerns, in the unitary gauge the most general contribution to the action from the inflationary sector is \cite{Cheung:2007st}
\begin{align} \label{unitaryinf} 
\nn  S_{g}=&\int d^4x\sqrt{-g}\Bigg[\frac{1}{2}m^2_{Pl}R+m^2_{Pl}\dot{H}g^{00}-m^2_{Pl}\left(3H^2+\dot{H}\right)\\
\nn &+\frac{M_2(t)^4}{2!}\left(g^{00}+1\right)^2
-\frac{\bar{M}_1(t)^3}{2}\left(g^{00}+1\right)\delta {K^\mu}_\mu-\frac{\bar{M}_2(t)^2}{2}(\delta{K^\mu}_\mu)^2\\
& +\frac{M_3(t)^4}{3!}\left(g^{00}+1\right)^3-\frac{\bar{M}_3(t)^2}{2}\delta{K^\mu}_\nu\delta{K^\nu}{_\mu}+...\Bigg] \end{align}
In the first line the coefficients of the linear terms are determined by demanding that variation of this action with respect to the metric give the Friedmann equations at zero order in perturbations. The coefficients of the rest of the terms, which are the higher order perturbation terms, are undetermined. All that can be said of them in general is that they are time dependent and their characteristics convey the properties of the background to the level of perturbations. These terms involve different effects, for instance $M_2$ introduces a sound speed for the inflaton perturbation, where as $\bar{M}_3$ modifies the speed of propagation for the tensor degrees of freedom. The action \eqref{unitaryinf} as it appears includes all terms that respect the symmetries within the unitary gauge. A specific model of interest will involve only some of these terms. 

The preheating field $\chi$ transforms as a scalar under the remaining possible spatial diffeomorphisms. It can be added to the system with the following terms \cite{Noumi:2012vr}
\begin{align}
S_\chi=\int d^4 x \sqrt{-g}\left[-\frac{\alpha_1(t)}{2}\partial^\mu\chi\partial_\mu\chi+\frac{\alpha_2(t)}{2}\left(\partial^0\chi\right)^2-\frac{\alpha_3(t)}{2}\chi^2+\alpha_4(t)\chi\partial^0\chi\right]
\end{align}
Here $\alpha_2$ introduces a sound speed in the preheating sector. Possible effects of $\alpha_2$ on sourcing secondary gravitational waves, and the effects of $\alpha_3$ for $\chi$ production through resonance were discussed in \cite{Giblin:2017qjp}. Via $\delta g_{\mu\nu}$, the terms above involve mixing between the inflationary and preheating sectors. In addition to these there can also be the following voluntary contributions \cite{Noumi:2012vr}  
\begin{align}
S_{g\chi}=\int d^4 x \sqrt{-g}\left[\beta_1(t)\delta g^{00}\chi+\beta_2(t)\delta g^{00}\partial^0\chi+\beta_3(t)\partial^0\chi-(\dot{\beta}_3(t)+3H(t)\beta_3(t))\chi\right].
\end{align}
Notice that these later contributions involve derivative couplings between the two sectors. It is these terms that we will focus on in the following.

A time diffeomorphism $t\to\tilde{t}=t+\xi^0$ performed to this action, requires the introduction of the scalar $\pi$ which transforms as $\pi\to\tilde{\pi}=\pi-\xi^0$, to guarantee that the parameter $\xi^0$ does not begin to appear in the action. The unitary gauge discussed above corresponds to the gauge where $\pi$ was set to zero. Via such transformations $\pi$ becomes explicit in the action. Following \cite{Noumi:2012vr}, as
\begin{align}
g^{00}&\to g^{00}+2g^{0\mu}\partial_\mu\pi+g^{\mu\nu}\partial_\mu\pi\partial_\nu\pi\\
\beta_i(t)&\to\beta_i(t+\pi)\\
\partial^0\chi&\to\partial^0\chi+g^{\mu\nu}\partial_\mu\chi\partial_\nu\pi\\
\int d^4x\sqrt{-g}&\to\int d^4x\sqrt{-g}.
\end{align}
the action for the inflationary sector becomes
\begin{align} \label{pigaugeinf}
\nn S_g=&\int d^4 x \sqrt{-g}\Bigg[\frac{1}{2}m^2_{Pl}R-m^2_{Pl}\left(3H^2(t+\pi)+\dot{H}(t+\pi)\right)+\\
\nn &+m^2_{Pl}\dot{H}(t+\pi)\left((1+\dot{\pi})^2g^{00}+2(1+\dot{\pi})\partial_i\pi g^{0i}+g^{ij}\partial_i\pi\partial_j\pi\right)\\
&+\frac{m_2(t+\pi)^4}{2!}\left((1+\dot{\pi})^2g^{00}+2(1+\dot{\pi})\partial_i\pi g^{0i}+g^{ij}\partial_i\pi\partial_j\pi+1\right)^2+...\Bigg],\end{align}
the reheating sector takes the form
\begin{align} S_\chi=\int d^4 x \sqrt{-g}\Bigg[&-\frac{\alpha_1(t+\pi)}{2}\partial^\mu\chi\partial_\mu\chi+\frac{\alpha_2(t+\pi)}{2}\left(\partial^0\chi+g^{\mu\nu}\partial_\mu\chi\partial_\nu\pi\right)^2\\
&-\frac{\alpha_3(t+\pi)}{2}\chi^2+\alpha_4(t+\pi)\chi\left(\partial^0\chi+g^{\mu\nu}\partial_\mu\chi\partial_\nu\pi\right)\Bigg]
\end{align}
and the action that describes further mixings at second order becomes
\begin{align}
& S^{(2)}_{g\chi}=\int d^4 x\sqrt{-g}\Bigg[ \beta_1(t+\pi)\left(\delta g^{00}+2\partial^0\pi+\partial_\mu\pi\partial^\mu\pi\right)\chi\\
&+\beta_2(t+\pi)\left(\delta g^{00}+2\partial^0\pi+\partial_\mu\pi\partial^\mu\pi\right)\left(\partial^0\chi+\partial_\nu\pi\partial^\nu\chi\right)\\
&+\beta_3(t+\pi)\left(\partial^0\chi+\partial_\mu\pi\partial^\mu\chi\right)-\left(\dot{\beta}_3(t+\pi)+3H(t+\pi)\beta_3(t+\pi)\right)\chi\Bigg].\end{align}

Starting from the unitary gauge one can move to the gauge in which inflationary perturbations appear explicitly as $\delta \phi$ through the diffeomorphism with parameter $\xi^0=-\pi$. The two variables $\delta\phi$ and $\pi$ both denote perturbations to the inflaton and are related to each other via
\be \delta\phi=\pi_c=\frac{\pi}{c_\pi}\sqrt{-2m^2_{pl}\dot{H}}=\frac{\pi}{c_\pi}\dot{\phi}_0.\ee
Even though the action is constructed by making a gauge choice first, the system on the whole always respects invariance under all of the diffeomorphisms. 

Time diffeomorphims are gauge symmetries. In the presence of backgrounds with time dependence, there is a divergent charge associated with them. This implies that the global time translation invariance is also not respected by the background. Given the isotropy and homogeneity of the background these diffeomorphisms can be thought of as approximate gauge or global symmetries. Such backgrounds, where cosmological backgrounds are included, are said to \emph{spontaneously break} time diffeomorphism invariance. The scale at which the charge associated with the global time translation invariance becomes ill defined sets the symmetry breaking scale \cite{Cheung:2007st, Baumann:2011su}
\be\label{symbr} \Lambda_{sb}^2=\sqrt{-2m^2_{pl}\dot{H}c_\pi},\ee
where $c^{-2}_\pi=1-\frac{M^4_2}{m^2_{pl}\dot{H}}$. 
Hence this EFT where scalar inflaton perturbations appear as Goldstone modes of nonlinearly realized time diffeomorphisms is valid below the scale $\Lambda_{sb}$. 
\section{Background and Scales for Preheating}
\label{sec:background scales}

If the energy momentum density of the universe is dominated by a single scalar matter field, the behavior of the Hubble parameter at the end of inflation, for times such that $1
\ll m_\phi t$ is
\be \label{Hpre} H_p=H_m(t)-\frac{3}{4}\frac{H^2_m}{m_\phi}sin(2m_\phi t),\ee
as summarized in appendix \ref{ap:background}. Here $m_\phi$ denotes the mass of the inflaton field, who remains to be the dominant matter source in the early stages of preheating. At the end of inflation, whatever may be the potential that drives inflation, the inflaton field oscillates at the minimum of its potential. By then, this potential can be approximated via the Taylor expansion as $V\sim\frac{1}{2}m_\phi^2\phi_0^2$. The mass of the inflaton sets the frequency of these oscillations, which start at times when the Hubble parameter drops below the mass of the inflaton $H_p(t)<m_\phi$. 

Hence during preheating there are two scales associated with the background, the mass of the inflaton field $m_\phi$ and the Hubble parameter $H_p$. And these scales possess the hierarchy $H_p<m_\phi$ between each other. This hierarchy guarantees that \eqref{Hpre} does not describe a background that oscillates, which would be problematic. Instead the preheating background resembles matter domination with oscillatory contributions which are suppressed by $\frac{H_m}{m_\phi}$.

The background \eqref{Hpre} fits a symmetry breaking pattern  
\be H(t)=H_{FRW}(t)+H_{osc}P(\omega t)\ee   
where $P(\omega t)$ can be any periodic function. This resembles the discrete symmetry breaking pattern described for inflation \cite{Behbahani:2011it}. While inflation corresponds to a weak breaking of time translation invariance and can respect discrete time translation invariance, the strong time dependence of $H_m$ implies that during preheating time translation invariance is completely broken. The time dependence of coefficients $H_{FRW}=H_m$ and $H_{osc}=-\frac{3}{4}\frac{H^2_m}{m_\phi}$ is not weak, but it is required that this time dependence is small compared to the scale of oscillations $\omega=2m_\phi$
\be \frac{\dot{H}}{\omega H}\sim\frac{\dot{H}_m}{\omega H_m}\sim\frac{\dot{H}_{osc}}{m_\phi H_{osc}}\ll 1.\ee
Thus the hierarchies between scales associated with the preheating background go as
\be \omega\gg H_p\sim H_m,~~H_m\gg H_{osc},~~\omega H_m\gg \dot{H}_m.\ee
These hierarchies imply the following form for the derivatives of the Hubble parameter\footnote{The dominant terms in these derivatives are\\ $\dot{H}_p=-\frac{3}{2}\left(1+cos(2m_\phi t+\Delta)\right)H^2_m,$\\
	$\ddot{H}_p=3m_\phi H^2_msin(2m_\phi t+\Delta),$\\
	$\dddot{H}_p=6m_\phi^2H^2_mcos(2m_\phi t+\Delta).$\\
	$\ddddot{H}_p=-12m_\phi^3H^2_{m}sin(2m_\phi t).$
	Since we are focusing on times $m_\phi t\gg1$, the oscillations are frequent enough for us to approximate $\ddot{H}_p$ and $\dddot{H}_p$ by their amplitudes. In the main text we also neglect the overall numerical factors in these amplitudes. Our main objective in the next sections will be to emphasize how the frequency of the oscillations $\omega=2m_\phi$ becomes explicit in the scales of the problem.}
\be \dot{H}_p\sim -H^2_p,~~\ddot{H}_p\sim\omega H^2_p,~~\dddot{H}_p\sim \omega^2H^2_p.\ee

While there is expansion at the background, particle production through energy transfer from the background to the reheating field should be a local process, which respects conservation of total energy density. It is a process that merely transfers energy from one species to another. This locality is achieved by focusing on the flat space limit $a\to1$. This limit implies $\dot{H}\to0$ and needs to be accompanied with the limit $m^2_{pl}\dot{H}\to finite$, to guarantee that the combination of scales that appear in the linear order action with dimensions of energy density remain finite. During preheating this limit also implies $m^2_{pl}H^2\to finite$ as $H\to 0$.

The $\alpha_i$ terms lead to particle production. In theory, all the terms of similar dimensions in $S_g$, $S_\chi$ or $S_{g\chi}$ are expected to be of equal strength. In practice, some of these terms are turned off in order to be able to focus on different effects one wishes to address. As the linear order action possess $m^2_{pl}\dot{H}\sim m^2_{pl}H^2$, where the similarity holds during preheating, it is natural to expect all the EFT parameters $\{\alpha_i,M_i,\beta_i\}$, to be proportional to $m^2_{pl}\dot{H}$. Since $m_\phi$ is the leading scale among the scales $m_\phi$ and $H_p$ associated with the background, it can be used to adjust for the dimensions of the EFT parameters. For example $\alpha_3$ has mass dimension $M^2$, while $\alpha_1$ has mass dimension zero. So these EFT parameters can be approximated as
\be \alpha_1\sim \frac{1}{m^4_\phi}m^2_{pl}\dot{H},\ee
while
\be \alpha_3\sim \frac{1}{m^2_{\phi}}m^2_{pl}\dot{H}.\ee
In general the equation of motion for the reheating perturbations is of the form
\be \ddot{\tilde\chi}_c+\omega_\chi^2(t)\tilde\chi_c=0,\ee
where the subscript and tilde denote that the field has been canonically normalized $\tilde{\chi}_c\equiv \sqrt{\alpha_1+\alpha_2}\chi a^{3/2}$. Similarly we also absorb a factor of $a^3$ by defining $\tilde{\pi_c}\equiv a^{3/2}\pi_c$, which in return guarantees that the equation of motion for inflaton perturbations will also be of the form of a canonically normalized harmonic oscillator. To ease notation in calculations that will follow, we also define ${\tilde{m}_\chi}^2\equiv m^2_\chi-\left(\frac{9}{4}H^2+\frac{3}{2}\dot{H}\right)$, and ${\tilde{m}_\chi}^2\equiv m^2_\chi-\left(\frac{9}{4}H^2+\frac{3}{2}\dot{H}\right)$. For canonical quantization, it will be the canonically normalized fields $\tilde{\pi}_c$, $\tilde{\chi}_c$ and their conjugate momenta who satisfy the canonical commutation relations.

 The frequency $\omega_\chi(t)$ will possess some time dependence coming from the time dependence of $\alpha_i(t)$.\footnote{For the canonically normalized field $\tilde{\chi}_c$, the couplings with the background induce the time dependent mass $m^2_\chi(t)=\frac{\alpha_3(t)-\dot{\alpha}_4(t)}{\alpha_1(t)+\alpha_2(t)}-\frac{1}{2}\frac{\ddot{\alpha}_1+\ddot{\alpha}_2}{\alpha_1+\alpha_2}+\frac{1}{4}\left(\frac{\dot{\alpha_1}+\dot{\alpha}_2}{\alpha_1+\alpha_2}\right)^2$. The frequency of $\tilde{\chi}_c$ modes are $\omega^2_k=c^2_\chi(t)\frac{k^2}{a^2}+\tilde{m}^2_\chi$.} Particle production occurs at times when the time dependence in $\omega_\chi(t)$ becomes nonadiabatic. This is independent from the adiabaticity of the background $H_p(t)$.  This time dependence, and hence $\chi$ production, can be sourced purely due to a time dependent sound speed in the reheating sector $c_\chi^2=\frac{\alpha_1}{\alpha_1+\alpha_2}$, or it can be sourced by the couplings between the reheating sector and the background. In the later case, particle production with constant sound speed occurs for modes in the range \cite{Giblin:2017qjp, Sengor:2018hkb}
\be \label{bckprod} k^2<\frac{1}{c^2_\chi}\left[\left(\frac{\dot{\alpha}_3-\ddot{\alpha}_4}{2}\right)^{2/3}-\frac{\alpha_3-\dot{\alpha}_4}{\alpha_1}\right]\equiv K^2_{bck}\propto \mathcal{O}\left(\frac{1}{c^2_\chi}\frac{m^2_{pl}H^2}{m^2_\phi}\right),\ee
which lies below the symmetry breaking scale. 
\section{Hidden Preheating}
\label{sec:Hidden}
Up to this point we have reviewed the background behavior and general form of interactions during preheating. Among the EFT parameters, $\alpha_i(t)$ control resonant $\chi_c$ production, while $H(t)$ and $m_i(t)$ affect the behavior of inflationary modes. The coefficients $\beta_i(t)$ denote additional couplings between the inflaton fluctuations and $\chi$. These kinds of interactions are unavoidable  if the inflaton and the reheat field are derivatively coupled to each other. In the unitary gauge, these interactions are
\be\label{hiddenS} S^c_{g\chi}=\int d^4x\sqrt{-g}\left[\beta_1(t)\delta g^{00}\chi+\beta_2(t)\delta g^{00}\partial^0\chi+\beta_3(t)\partial^0\chi-(\dot{\beta}_3(t)+3H(t)\beta_3(t))\chi\right],\ee
with the mass dimensions of the parameters being $[\beta_1]=M^3$ and $[\beta_2]=[\beta_3]=M^2$. In the following, these interactions will be the focus of our attention and we will proceed in the $\pi$-gauge with 
\begin{align}
\label{hiddenS1}& S^{(2)}_{g\chi}=\int d^4 x\sqrt{-g}\Bigg[ \beta_1(t+\pi)\left(\delta g^{00}+2\partial^0\pi+\partial_\mu\pi\partial^\mu\pi\right)\chi\\
\label{hiddenS2}&+\beta_2(t+\pi)\left(\delta g^{00}+2\partial^0\pi+\partial_\mu\pi\partial^\mu\pi\right)\left(\partial^0\chi+\partial_\nu\pi\partial^\nu\chi\right)\\
\label{hiddenS3}&+\beta_3(t+\pi)\left(\partial^0\chi+\partial_\mu\pi\partial^\mu\chi\right)-\left(\dot{\beta}_3(t+\pi)+3H(t+\pi)\beta_3(t+\pi)\right)\chi\Bigg].\end{align}
Let's begin by focusing on the renormalizable quadratic couplings
\begin{align} S^{(2)}_{g\chi}\supset\int& d^4xa^3\Bigg[\beta_1(t)\left(\delta g^{00}\chi-2\dot{\pi}\chi\right)+\beta_2(t)\left(-\delta g^{00}\dot{\chi}+2\dot{\pi}\dot{\chi}\right)\\
&-\delta N\beta_3(t)\dot{\chi}-\dot{\beta}_3(t)\pi\dot{\chi}+\beta_3(t)\delta g^{0\mu}\partial_\mu\chi-\beta_3(t)\dot{\chi}\dot{\pi}\\
&+\beta_3(t)\partial_i\chi\partial_i\pi-\ddot{\beta}_3(t)\pi\chi-(3\dot{\beta}_3H+3\beta_3\dot{H})\pi\chi\Bigg],\end{align}
and keeping in mind the connection
\be \delta\phi=\pi_c=\frac{\pi}{c_\pi}\sqrt{-2m^2_{pl}\dot{H}}=\frac{\pi}{c_\pi}\dot{\phi}_0.\ee
Consider the terms with temporal derivatives. After canonical normalization with $\alpha_1+\alpha_2=1$, and $\delta g^{00}=\frac{\delta g^{00}_c}{m_{pl}}=\frac{\delta N_c}{m_{pl}}$ these give\footnote{For reference, in the case where $m_i=\beta_i=0$ which would mean no derivative coupling, the constraints give $\delta N=-\frac{\dot{H}}{H}\pi$. This makes $\delta N_c=m_{pl}\delta N=\frac{\sqrt{-\dot{H}}}{\sqrt{2}H}\pi_c\sim\frac{\pi_c}{\sqrt{2}}$.} 
\begin{align}
S^{(2)}_{g\chi}&\supset\int d^4 xa^3\Bigg[\beta_1\delta N_c\frac{\chi_c}{m_{pl}}-2\beta_1\frac{c_\pi}{\sqrt{-2m^2_{pl}\dot{H}}}\left(\frac{\dot{c}_\pi}{c_\pi}\pi_c+\dot{\pi}_c-\frac{\ddot{H}}{2\dot{H}}\pi_c\right)\chi_c\\
&-\beta_2\frac{\delta N_c}{m_{pl}}\dot{\chi}_c+2\beta_2\frac{c_\pi}{\sqrt{-2m^2_{pl}\dot{H}}}\left(\frac{\dot{c}_\pi}{c_\pi}\pi_c+\dot{\pi}_c-\frac{\ddot{H}}{2\dot{H}}\pi_c\right)\dot{\chi}_c+\beta_3\frac{\delta g^{0i}_c}{m_{pl}}\partial_i{\chi}_c\\
&-\frac{c_\pi}{\sqrt{-2m^2_{pl}\dot{H}}}\dot{\beta}_3\dot{\chi}_c\pi_c-\beta_3\frac{c_\pi}{\sqrt{-2m^2_{pl}\dot{H}}}\left[-\frac{1}{2}\frac{\ddot{H}}{\dot{H}}\pi_c+\dot{\pi}_c+\frac{\dot{c}_\pi}{c_\pi}\pi_c\right]\dot{\chi}_c\\
&~~~~~~~~~~-\frac{c_\pi}{\sqrt{-2m^2_{pl}\dot{H}}}\ddot{\beta}_3\pi_c\chi_c-\frac{c_\pi}{\sqrt{-2m^2_{pl}\dot{H}}}(3\dot{\beta}_3H+3\beta_3\dot{H})\pi_c\chi_c\Bigg]\end{align}
Our study of the background taught us that higher derivatives on $H$ are stronger because they involve more powers of $m_\phi$. Derivatives of $H$ appear in $S^{(2)}$ after canonical normalization. Looking at all the terms in $S^{(2)}$, the terms that involve the most number of derivatives will be stronger among the terms of same order in perturbations.\footnote{This does not imply that the next order action will be stronger then the previous. For example at third order one has the term $S^{(3)}_{g\chi}\supset \gamma \ddot{\pi}\chi\dot{\chi}$ \cite{Noumi:2012vr}. This will involve $\gamma\left(-\frac{\dddot{H}}{\dot{H}}+\frac{\ddot{H}^2}{\dot{H}^2}\right)\frac{\dot{\pi}_c}{\sqrt{-2m^2_{pl}\dot{H}}}\chi_c\dot{\chi}_c\sim\gamma \frac{m^2_\phi\omega_k}{m_{pl}H}\pi_c\chi_c^2$where $\gamma$ is dimension zero and this term is highly suppressed via the symmetry breaking scale, compared to terms in $S^{(2)}$.} For each parametrization the strongest terms are
\begin{align}
\beta_1:&\beta_1\frac{c_\pi}{\sqrt{-2m^2_{pl}\dot{H}}}\frac{\ddot{H}}{\dot{H}}\pi_c\chi_c~\text{and}~-2\beta_1\frac{c_\pi}{\sqrt{-2m^2_{pl}\dot{H}}}\dot\pi_c{\chi}_c\\
\beta_2:&-\beta_2\frac{c_\pi}{\sqrt{-2m^2_{pl}\dot{H}}}\frac{\ddot{H}}{\dot{H}}\pi_c\dot{\chi}_c~\text{and}~2\beta_2\frac{c_\pi}{\sqrt{-2m^2_{pl}\dot{H}}}\dot\pi_c\dot{\chi}_c\\
\beta_3:&-\frac{c_\pi\dot{\beta}_3}{\sqrt{-2m^2_{pl}\dot{H}}}\dot{\chi}_c\pi_c,~ -\frac{\beta_3 c_\pi}{\sqrt{-2m_{pl}\dot{H}}}\frac{\ddot{H}}{\dot{H}}\pi_c\dot{\chi}_c,~ -\frac{c_\pi\ddot{\beta}_3}{\sqrt{-2m^2_{pl}\dot{H}}}\pi_c\chi_c.
\end{align}

Among these terms, there are the derivative couplings\footnote{The terms coming from metric perturbations via $\delta N_c$ can be ignored as they are Planck Suppressed at this order.} 
\be \frac{\beta_1}{\sqrt{-2m^2_{pl}\dot{H}}}\dot{\pi}_c\chi_c,~~\frac{\beta_2}{\sqrt{-2m^2_{pl}\dot{H}}}\frac{\ddot{H}}{\dot{H}}\pi_c\dot{\chi}_c,~~\frac{\beta_3}{\sqrt{-2m^2_{pl}\dot{H}}}\pi_c\dot{\chi}_c,\ee
which are of the form $R_1 \dot{\pi}_c\chi_c$ and $R_3\pi_c\dot{\chi}_c$.
These terms can compete with the kinetic terms. Notice that $R_{1,3}$ has dimensions of energy, so it sets the energy scale for these derivative interactions. Below the scale set by $R_{1,3}$, if these derivative couplings dominate over the kinetic terms $\dot{\pi}_c^2$, $\dot{\chi}_c^2$, the system will effectively have a single degree of freedom. This is the same situation that is addressed in \cite{Baumann:2011su}, during inflation. Following the same line of thought, let us consider the consequences during preheating. For the moment the sound speeds $c_\chi$, $c_\pi$ can be set to unity, since how they can amplify or reduce the strength of these interactions is not the main concern. This amounts to setting $m_2=0$, $\alpha_1=1$.

To see how the number of effective degrees of freedom goes down to being single, let us consider the quadratic Lagrangian for the canonically normalized perturbations in the presence of these interactions one by one. The scales involved will be considered within the limit $a\to 1$, keeping $m^2_{pl}\dot{H}$ finite, which was also meaningful for particle production purposes. 
\subsection{Hidden Preheating by $\beta_1(t)$}
\label{sec:beta1}

In the presence of $\beta_1$ the Lagrangian becomes,
\begin{align}
\nn  L^{(2)}=\int d^3 xa^3\mathcal{L}=\int d^3x a^3\Big[ \frac{1}{2}\dot{\chi}_c^2+\frac{1}{2}\dot{\pi}_c^2-\frac{1}{2a^2}(\partial_i\chi_c)^2-\frac{1}{2a^2}(\partial_i\pi_c)^2\\
-\frac{1}{2}m^2_\chi\chi_c^2-\frac{1}{2}m^2_\pi\pi_c^2-2R_1\dot{\pi}_c\chi_c-R_1\frac{\ddot{H}}{\dot{H}}\pi_c\chi_c\Big]\end{align}
where $R_1\equiv\frac{\beta_1(t)}{\sqrt{-2m^2_{pl}\dot{H}}}$. In terms of the canonically normalized fields this Lagrangian is
\begin{align}\nn L=\int d^3x\frac{1}{2}\Big[\dot{\tilde{\pi}}_c^2&-\tilde{m}_\pi^2\tilde{\pi}_c^2-\frac{1}{a^2}\left(\partial_i\tilde{\pi}_c\right)^2+\dot{\tilde{\chi}}_c^2-\tilde{m}_\chi^2\tilde{\chi}_c^2-\frac{1}{a^2}\left(\partial_i\tilde{\chi}_c\right)^2\\
&-4R_1\dot{\tilde{\pi}}_c\tilde{\chi}_c+6HR_1\tilde{\pi}_c\tilde{\chi}_c-2R_1\frac{\ddot{H}}{\dot{H}}\tilde{\pi}_c\tilde{\chi}_c\Big]\end{align}

 The equations of motion for modes of each species are as follows
\begin{align}\label{chieom_beta1}
\ddot{\tilde\chi}_{ck}+\left[\frac{k^2}{a^2}+\tilde{m}_\chi^2(t)\right]\tilde{\chi}_{ck}&=R_1\left[-2\dot{\tilde\pi}_{kc}+\left(3H-\frac{\ddot{H}}{\dot{H}}\right)\tilde{\pi}_{ck}\right],\\
\label{pieom_beta1}\ddot{\tilde\pi}_{ck}+\left[\frac{k^2}{a^2}+\tilde{m}^2_\pi(t)\right]\tilde{\pi}_{ck}&=R_1\left[2\dot{\tilde{\chi}}_{ck}+\left(2\frac{\dot{R}_1}{R_1}+3H-\frac{\ddot{H}}{\dot{H}}\right)\tilde{\chi}_{ck}\right].
\end{align}
From the right hand sides of equations \eqref{chieom_beta1} and \eqref{pieom_beta1} one can see the source terms introduced for each species because of the $\beta_1$ coupling. We are interested in analyzing how these terms modify the dispersion relation in general. In a different direction, if one is interested in considering particle production in the later stages of preheating, then these terms become important for capturing back reaction effects between the two sectors. 

 While WKB-like solutions $\tilde{\pi}_{ck}\sim e^{-i\int\omega(t')dt'}$, $\tilde{\chi}_{ck}\sim e^{-i\int\omega(t')dt'}$ hold, the kinetic terms go by $\omega^2$, and the kinetic coupling goes as $R_1\omega$ . In the range $R_1\gg \omega$, the kinetic coupling dominates over the kinetic terms, the kinetic terms are negligible and the theory can be approximated by 
\begin{align} \label{hiddenbeta1lag} \nn L\simeq\int d^3x\Big[ -2R_1\dot{\tilde\pi}_c\tilde{\chi}_c-\frac{1}{2a^2}(\partial_i\tilde{\chi}_c)^2-\frac{1}{2a^2}(\partial_i\tilde{\pi}_c)^2\\
-\frac{1}{2}{\tilde{m}}^2_\chi\tilde{\chi}_c^2-\frac{1}{2}{\tilde{m}}^2_\pi\tilde{\pi}_c^2+3HR_1\tilde{\pi}_c\tilde{\chi}_c-\frac{\ddot{H}}{\dot{H}}R_1\tilde{\pi}_c\tilde{\chi}_c\Big].\end{align}
At first sight it looks like the kinetic term here has a wrong sign, but this depends on the sign of $R_1$, which is not necessarily positive, as will be demonstrated below with a specific example. 

In this range $\tilde{\chi}_c$ is no longer a dynamical field, since it doesn't have any kinetic terms and its corresponding conjugate momentum is constrained to vanish. Instead it plays the role of canonical momenta for $\tilde{\pi}_c$,
\be\label{conjmom_beta1} p_\pi\equiv\frac{\partial\mathcal{L}}{\partial\dot{\tilde\pi}_c}=-2R_1\tilde{\chi}_c,~~~p_\chi\equiv\frac{\partial\mathcal{L}}{\partial\dot{\tilde{\chi}}_c}=0.\ee



The original motivation for introducing the reheating field was to populate the universe with its perturbations. One would have liked to see more and more of $\tilde{\chi}_c$ modes being the effective degrees of freedom during preheating, yet it turns out that in the presence of $\beta_1$ interactions, for scales $R_1\gg \omega$ it is only the inflaton perturbations who propagate as the effective degrees of freedom! This is not to say that the presence of the reheat perturbations go completely unnoticed. They affect the system by determining the canonical momenta of $\tilde{\pi}_c$. So in a sense this is a type of reheating where there is a range of energies in which the reheating field determines the dynamics of the inflaton perturbations, while it itself stays hidden. For this reason let us refer to this regime as the regime of ``Hidden Preheating''. 

Having noticed this Hidden Preheating regime, now is a good point to analyze the dispersion relation in this regime. To make matters more simple for this purpose, let us leave aside the resonant particle production effects by dropping the last two terms in \eqref{hiddenbeta1lag} and assume $R_1$, $\tilde{m}_\pi$, $\tilde{m}_\chi$ do not have strong time dependence. These assumptions can be interpreted to mean that we are considering scales such that $R_1 \gg \omega \gg m_\phi\gg H_p$ and neglecting the time dependence of the EFT parameters $\alpha_i(t)$ and $\beta_1(t)$, which for practical purposes amounts to taking $a\to 1$.

Vanishing of $p_\chi$ among the canonical momenta is a primary constraint of the system. It means that not all of the canonical variables, $\{\tilde{\pi}_c, p_\pi, \tilde{\chi}_c, p_\chi\}$, are physical degrees of freedom. Some of them are redundant variables that can be set to zero. For the consistency of the system this constraint must be preserved with time, that is the condition
\be \label{consistcond} \dot{p}_\chi=0\ee
should be satisfied. With the above assumptions the Hamiltonian corresponding to \eqref{hiddenbeta1lag} is
\begin{align}\label{hiddenbeta1ham}
\nn H=&\int d^3x\left[p_\pi\dot{\tilde{\pi}}_c+p_\chi\dot{\tilde{\chi}}_c\right]-L\\
=&\frac{1}{2}\int d^3x\left[\left(\partial_i\tilde{\pi}_c\right)^2+\frac{1}{4R^2_1}\left(\partial_ip_\pi\right)^2+\tilde{m}_\pi^2\tilde{\pi}_c^2+\frac{\tilde{m}_\chi^2}{4R^2_1}p^2_\pi\right]\end{align}
where equations \eqref{conjmom_beta1} have been employed in the last line. By the Hamilton equations of motion
\be \dot{p}_\chi=-\frac{\partial H}{\partial\tilde{\chi}_c}=0\ee
and thus the consistency condition \eqref{consistcond} is satisfied. Once the fields are decomposed into Fourier modes, the equations of motion for the remaining variables are
\be \label{hameomchi_beta1} \dot{\tilde{\chi}}_{ck}=\frac{\partial H}{\partial p_{\chi k}}=0,\ee
\be \label{hameomppi_beta1} \dot{p}_{\pi k}=-\frac{\partial H}{\partial \tilde{\pi}_{ck}}=-\left[k^2+\tilde{m}_\pi^2\right]\tilde{\pi}_{ck},\ee
\be \label{hameompi_beta1} \dot{\tilde{\pi}}_{ck}=\frac{\partial H}{\partial p_{\pi k}}=\frac{1}{4R^2_1}\left[k^2+\tilde{m}_\chi^2\right]p_{\pi k}.\ee
 
Equations \eqref{hameompi_beta1} and \eqref{hameomppi_beta1} make up a coupled system of differential equations. By differentiating \eqref{hameomppi_beta1} and employing \eqref{hameompi_beta1} one obtains

\be \label{diff_pi_beta1} \ddot{\tilde{\pi}}_{ck}=-\frac{1}{4R^2_1}\Big[k^4+\left(\tilde{m}_\pi^2+\tilde{m}_\chi^2\right)k^2+\tilde{m}_\pi^2\tilde{m}_\chi^2\Big]\tilde{\pi}_{ck}.\ee
This has solutions of the form $\tilde{\pi}_{ck}\sim e^{-i\omega t}$ with
\be \label{omega_hiddenregion1} \omega=\frac{1}{2R_1}\sqrt{k^4+\left(\tilde{m}_\pi^2+\tilde{m}_\chi^2\right)k^2+\tilde{m}_\pi^2\tilde{m}_\chi^2}\ee
at scales $R_1\gg \omega\gg m_\phi>H_p$.

In summary, in the Hidden Preheating regime of $\beta_1$ coupling, the effective degrees of freedom that propagate are the $\tilde{\pi}_{ck}$ modes with the dispersion relation \eqref{omega_hiddenregion1}. In this regime the $\tilde{\chi}_c$ sector has been integrated out. However the present constraints on the $\tilde{\chi}_c$ must be treated with care during quantization in this regime. Similar to the situation discussed in \cite{Baumann:2011su} during inflation, the hidden presence of the reheating perturbations leads to a sound speed and further modifications in the dispersion relation of the inflaton perturbations. For modes in the range $R_1\gg k>\tilde{m}_\pi,\tilde{m}_\chi$ this modification at leading order is
\be \label{hiddenbeta1_rkm} \omega \simeq \frac{k^2}{2R_1}.\ee

And for the longest wavelength modes in the range $R_1\gg \tilde{m}_\pi,\tilde{m}_\chi >k$ the modification becomes
\begin{align} \nn \omega&\simeq \frac{\tilde{m}_\chi\tilde{m}_\pi}{2R_1}\sqrt{1+\frac{\tilde{m}_\pi^2+\tilde{m}_\chi^2}{\tilde{m}_\pi^2\tilde{m}_\chi^2}k^2}\\
&\simeq \frac{\tilde{m}_\pi \tilde{m}_\chi}{2R_1}+\frac{\tilde{m}_\pi^2+\tilde{m}_\chi^2}{\tilde{m}_\pi^2\tilde{m}_\chi^2}\frac{k^2}{4R_1}+...\end{align}
 
Outside of the Hidden Preheating regime, with the ansatz $\tilde{\chi}_{ck}\sim Ae^{-i\omega t}$, $\tilde{\pi}_{ck}\sim Be^{-i\omega t}$ and our previous assumptions on neglecting the time dependence of the background and EFT coefficients, equations \eqref{chieom_beta1} and \eqref{pieom_beta1} lead to the following dispersion relation
\be \omega_\pm^2=k^2+\frac{\tilde{m}^2_\chi+\tilde{m}^2_\pi}{2}+2R_1^2\pm \sqrt{4R_1^2 k^2+\left(\frac{\tilde{m}^2_\chi+\tilde{m}^2_\pi}{2}+2R^2_1\right)^2-\tilde{m}^2_\pi \tilde{m}^2_\chi}.\ee
From here it is seen that in the range $k^2\gg4R_1^2,m^2_\chi,m^2_\pi$, modes of both the inflationary and reheating fields propagate independently with $\omega \sim k$.

For scales where $k$ and $R_1$ are comparable, by defining
\be \label{defA}\tilde{\mathcal{A}}^2\equiv\left(\frac{\tilde{m}^2_\chi+\tilde{m}^2_\pi}{2}+2R^2_1\right)^2-\tilde{m}^2_\pi \tilde{m}^2_\chi,\ee
we see that the modification gives rise to a sound speed and nonlinearities as follows
\be \omega^2_\pm\simeq k^2+\tilde{\mathcal{A}}\pm\left[\tilde{\mathcal{A}}+\frac{2R^2_1}{\tilde{\mathcal{A}}}k^2-\frac{2R^4_1}{\tilde{\mathcal{A}}^3}k^4+...\right].\ee
One of the solutions 
\be \label{omega+}\omega^2_+\simeq 2\tilde{\mathcal{A}}+\left(1+\frac{2R_1^2}{\tilde{\mathcal{A}}}\right)k^2\equiv 2\tilde{\mathcal{A}}+c^2_-k^2\ee
 describes very heavy modes that at leading order do not propagate and whose dispersion relation involves a sound speed $c^2_-$ in the next order contribution.\footnote{Note the sign difference in the definition of the sound speeds, $c^2_-=1+\frac{2R_1^2}{\tilde{\mathcal{A}}}$ and $c^2_+=1-\frac{2R^2_1}{\tilde{\mathcal{A}}}$. For causality the speed of propagation should not exceed the speed of propagation for light. In the units we are working with this reads $c^2_{\pm}\leq 1$. From equation \eqref{defA} $\tilde{\mathcal{A}}$ can have two different signs $\tilde{\mathcal{A}}=\pm\left[\left(\frac{\tilde{m}^2_\chi+\tilde{m}^2_\pi}{2}+2R^2_1\right)^2-\tilde{m}^2_\pi \tilde{m}^2_\chi\right]^{1/2}$. To satisfy the causality condition the positive solution for $\tilde{\mathcal{A}}$ must contribute to $c_+^2$ and the negative solution to $c_-^2$. The notation for sound speeds in equations \eqref{omega+} and \eqref{omega-} aim to emphasize this point.} The $\omega_-$ solution describes the light modes that obtain a sound speed starting from the leading order 
\be\label{omega-} \omega^2_-=\left(1-\frac{2R^2_1}{\tilde{\mathcal{A}}}\right)k^2+\frac{2R^4_1}{\tilde{\mathcal{A}}^3}k^4\equiv c_+^2k^2+\frac{2R_1^4}{\tilde{\mathcal{A}}^3}k^4.\ee
For scales in the range $c^2_+\frac{\tilde{\mathcal{A}}^3}{2R^4_1}<k^2<4R^2_1$, the second term in \eqref{omega-} will dominate over the first and at leading order the dispersion relation will be 

\be \label{omega-smallest}\omega_-^2\simeq \frac{2R_1^2}{\tilde{\mathcal{A}}^3}k^4~~~ \text{for}~~c^2_+\frac{\tilde{\mathcal{A}}^3}{2R^4_1}<k^2<4R^2_1.\ee 

The range $R_1\gg k> \tilde{m}_\chi,\tilde{m}_\pi$ considered in the previous analysis of the dispersion relation during the Hidden Preheating regime, falls within the range $c^2_+\frac{\tilde{\mathcal{A}}^3}{2R^4_1}<k^2<4R^2_1$ where equation \eqref{omega-smallest} holds. This range implies $\tilde{\mathcal{A}}^2\simeq4R^4_1$, which guarantees that expression \eqref{omega-smallest} matches with expression \eqref{hiddenbeta1_rkm}, and $c^2_+\sim 0$. Thus we know that the modes described by \eqref{omega-smallest} are the $\tilde{\pi}_{ck}$ modes in the Hidden preheating regime, while equations \eqref{omega+} and \eqref{omega-} describe both $\tilde{\chi}_{ck}$ and $\tilde{\pi}_{ck}$ modes.  


Let us try to make an estimate on the likeliness of such a range occurring, by making assumptions on the form of the EFT coefficients. As noted earlier, the mass dimensions of $\beta_i$ are $[\beta_1]=M^3$, $[\beta_2]=[\beta_3]=M^2$. Since the quadratic terms that were determined by the background are at order $m^2_{pl}\dot{H}$, $\beta_i$ is expected to have a similar form. Unless more is known about the background, $m_\phi$ which is the highest scale in the background evolution which can be used to make the dimensions fit
\be\label{convbeta1}\beta_1=b_1\frac{m^2_{pl}\dot{H}}{m_\phi},\ee \be\label{convbeta2}\beta_2=b_2\frac{m^2_{pl}\dot{H}}{m^2_\phi},\ee
\be\label{convbeta3}\beta_3(t)=b_3\frac{m^2_{pl}\dot{H}}{m^2_\phi}.\ee
As such
\be\label{rho1est} R_1\equiv\frac{\beta_1}{\sqrt{-2m^2_{pl}\dot{H}}}=\mathcal{O}\left(\frac{m_{pl}H}{m_\phi}\right)=\mathcal{O}\left(\frac{\Lambda^2_{sb}}{m_\phi}\right).\ee
So while in the range $\Lambda_{sb}>\omega_{\pi,\chi}>R_1$ there are 2 effective degrees of freedom $\pi_c$ and $\chi_c$, where as in the range $R_1>\omega_{\pi,\chi}$, the inflaton perturbations $\pi_c$ are the only effective degree. 

\begin{figure}[h]\label{fig:beta1modes}
	\includegraphics[width=\textwidth]{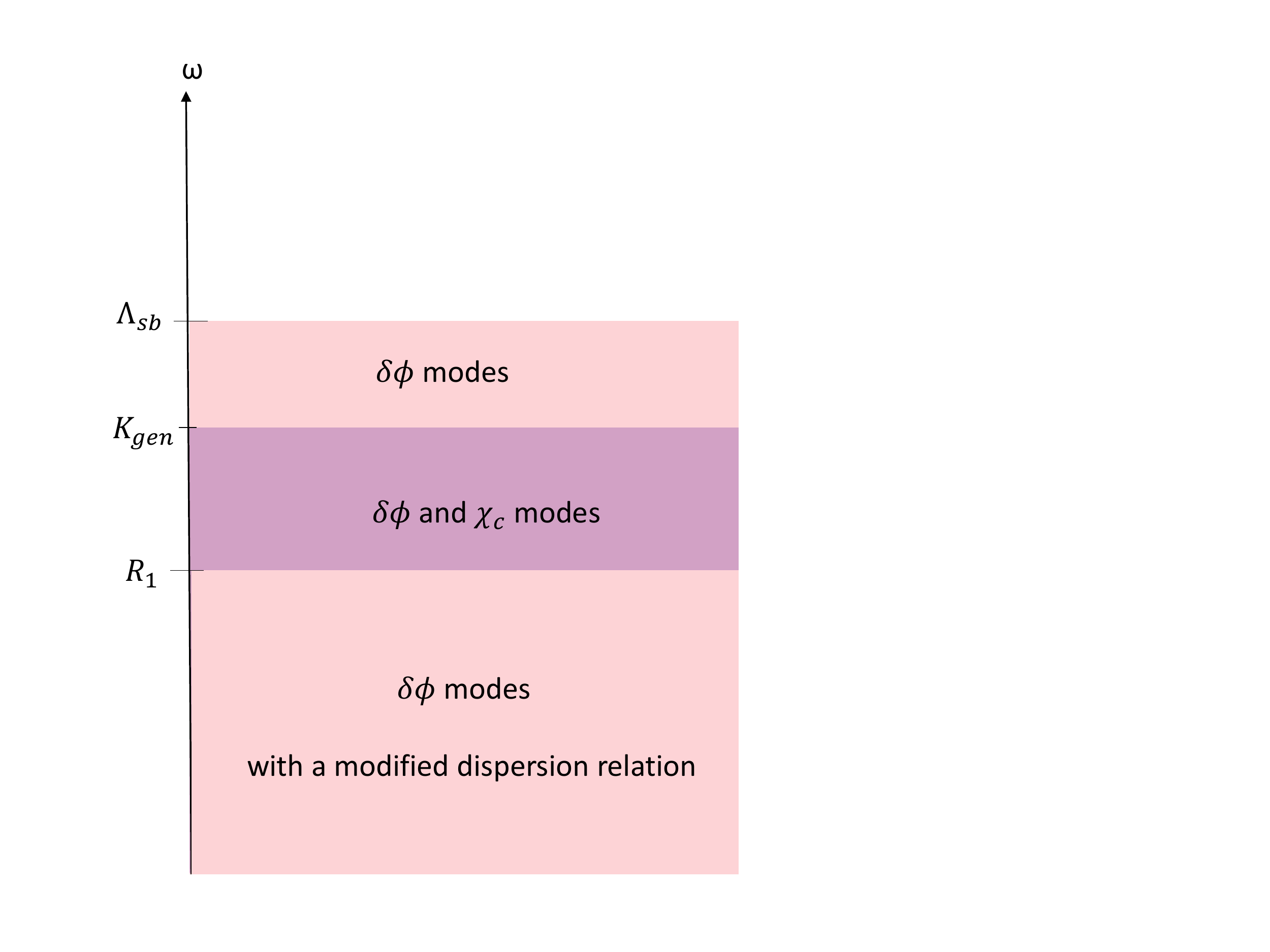}
	\caption{Here it is shown up to which scales the inflaton modes $\delta\phi=\pi_c$ and reheating modes $\chi_c$ appear as effective degrees of freedom, in the presence of $\beta_1$ type couplings.}
\end{figure}

Now we need to be a bit careful, does this leave any range for $\chi_c$-production? We found that the $\chi_c$ particles can be produced up to the scale
\be K^2_{bck}=\frac{1}{c_\chi^2}\left[\left(\frac{\dot{\alpha}_3-\ddot{\alpha}_4}{2}\right)^{2/3}-\alpha_3+\dot{\alpha}_4\right]\sim\mathcal{O}\left(\left[\frac{m^2_{pl}H^2}{m_\phi}\right]^{2/3}\right)=m_\phi^{2/3}R_1^{4/3}\ee
Considering \eqref{rho1est}, this suggests that the scale of particle production lies above the scale $R_1$, and so in the range $R_1<E<K_{bck}$ both $\pi_c$ and $\chi_c$ modes are effective degrees of freedom. We sumarize this distribution of effective modes with respect to scale in figure 1.

These $\beta_1$ interactions are present in models where the inflaton is derivatively coupled to the reheating sector. In the context of preheating derivative couplings have first been studied in \cite{ArmendarizPicon:2007iv}, with
\be\mathcal{L}=-\frac{1}{2}\partial^\mu\phi\partial_\mu\phi-\frac{1}{2}\partial^\mu X\partial_\mu X-V(\phi)- U(X)-\frac{1}{F^2}(\partial_\mu\phi\partial^\mu\phi)X^2\ee
where $F$ is the cutoff for this effective field theory. In the original work, the authors consider Chaotic Inflation in particular, to govern the inflaton sector in this low energy theory, in which case $F\simeq m_{pl}$. Another example is the case of geometric destabilization of inflation \cite{Renaux-Petel:2015mga}.

The Friedman equations at the background level are
\be -2m^2_{pl}\dot{H}=\left(1+2\frac{\chi_0^2}{F^2}\right)\dot{\phi}_0^2\equiv R^2\dot{\phi}_0^2\ee
\be 3m^2_{pl}H^2=\frac{1}{2}\left(1+2\frac{\chi_0^2}{F^2}\right)\dot{\phi}_0^2+V(\phi_0)+U(\chi_0)\ee
where $\chi_0$ is set to be constant. In general, such effective Lagrangians can also involve $-\frac{\partial_\mu\phi\partial^\mu\phi}{F}\chi$ terms, which would lead to $\chi_0(t)$ and give rise to an unstable growth in the reheating sector.

In unitary gauge, the fields are expanded in terms of linear perturbations as $\phi=\phi_0(t)$, $X=\chi_0+\chi(\vec{x},t)$ and, $g^{\mu\nu}=\bar{g}^{\mu\nu}(t)+\delta g^{\mu\nu}(\vec{x},t)$. With $V''(\phi_0)=m^2_\phi$ and $U''(\chi_0)=m^2_\chi$ the Lagrangian up to second order in perturbations is
\begin{align} \mathcal{L}^{(2)}&=-m^2_{pl}(3H^2(t)+\dot{H}(t))+m^2_{pl}\dot{H}g^{00}\\
&-\frac{1}{2}\partial_\mu\chi\partial^\mu\chi-\frac{1}{2}\left(m^2_\chi-2\frac{\dot{\phi}_0^2}{F^2}\right)\chi^2-2\frac{\chi_0}{F^2}\dot{\phi}^2\delta g^{00}\chi\end{align}
where the background equations of motion have been taken into account. This matches the EFT Lagrangian as a model with 
\be \alpha_1=1,~~\alpha_2=0,~~\alpha_3=m^2_\chi-2\frac{\dot{\phi}_0}{F^2}=m_\chi^2+\frac{4m^2_{pl}\dot{H}}{F^2},~~\alpha_4=0,~~m_i=0.\ee
Comparing the last line with \eqref{hiddenS} we can also read off that
\be \beta_1(t)=-2\frac{\chi_0}{F^2}\dot{\phi}_0^2=4\frac{\chi_0}{F^2R^2}m^2_{pl}\dot{H}.\ee
This sets the scale for Hidden preheating to be
\be R_1=\frac{\beta_1}{\sqrt{-2m^2_{pl}\dot{H}}}=-2\frac{\chi_0}{F^2R^2}\sqrt{-2m^2_{pl}\dot{H}}\sim\frac{\chi_0}{F^2R^2}\Lambda_{sb}^2.\ee
Also note that in the range $R_1>\omega$, the canonical momentum of the effective degree of freedom $\pi_c$ is
\be p_\pi=-2R_1\chi_c=4\frac{\chi_0}{F^2R^2}\sqrt{-2m^2_{pl}\dot{H}}~\chi_c.\ee

The $\chi_c$ production scale here is
\be K^2_{bck}=\left(\frac{2m^2_{pl}\ddot{H}}{F^2}\right)^{2/3}-m^2_\chi-4\frac{m^2_{pl}}{F^2}\dot{H}\ee
\be K^2_{bck}\sim\left(m_\phi\frac{\Lambda_{sb}^4}{F^2}\right)^{2/3}-m_\chi^2-4\frac{\Lambda_{sb}^4}{F^2}\ee
Corrections to $\phi$ dynamics here come with coefficients of $\frac{\chi_0}{F}$, which makes them perturbative corrections as long as $\chi_0\ll F$. This in return implies that $\frac{\chi_0}{R^2}=\frac{\chi_0}{\left(1+2\frac{\chi_0^2}{F^2}\right)^2}\sim\chi_0$, and the particle production scale will lie above the coupling $R_1^2\sim\frac{\chi_0^2}{F^2}\frac{\Lambda_{sb}^4}{F^2}$.

The derivative couplings preserve the shift symmetry of the inflaton. Hence they provide a very likely candidate for couplings of the inflaton with other fields. This also makes them more likely to be present in the later stages then nonderivative couplings, such as the original $g^2\phi^2X^2$ interaction considered for preheating. However, previous analysis of the instability regimes they lead to suggested that they are not very efficient for preheating. The line of inquiry here is showing that these type of derivative couplings lead to the presence of only inflationary perturbations with modified dispersion relations as the lightest modes present. This suggests a reason as to why they are inefficient for setting resonance in the reheating sector.
\subsection{Hidden Preheating with $\beta_2(t)$}
\label{sec:beta2}

The quadratic Lagrangian in the presence of $\beta_2$ is  
\be \mathcal{L}^{(2)}=\frac{1}{2}\dot{\chi}_c^2+\frac{1}{2}\dot{\pi}_c^2-\frac{1}{2a^2}(\partial_i\chi_c)^2-\frac{1}{2a^2}(\partial_i\pi_c)^2-\frac{1}{2}m^2_\chi\chi_c^2-\frac{1}{2}m^2_\pi\pi_c^2+2\rho_2\dot{\pi}_c\dot\chi_c-\rho_2\frac{\ddot{H}}{\dot{H}}\pi_c\dot\chi_c\ee
where $\rho_2\equiv\frac{\beta_2}{\sqrt{-2m^2_{pl}\dot{H}}}$. In the previous case, the coupling strength $R_1$ had mass dimension one and hence it defined a scale, but $\rho_2$ here is dimensionless. Moreover different then the case with $\beta_1$, here $\chi_c$ appears with time derivatives and there are two derivative couplings. 
Neglecting the background expansion and considering solutions of the form $\chi_c\sim Ae^{-i\omega t}$, $\pi_c Be^{-i\omega t}$ the strength of the kinetic terms will be of the order
\begin{align} \rho_2\dot{\chi}_c\dot{\pi}_c\sim \rho_2\omega^2 AB,~~\dot{\chi}_c^2\sim\omega^2A^2,~~\dot{\pi}_c^2\sim\omega^2B^2.\end{align} 
Being dimensionless $\rho_2$ does not define a scale and it will at most be order one. But if at times the amplitude of one of the species dominates over the other, the coupling with $\rho_2$ can dominate over the kinetic term for the species with the smaller amplitude and hence give it a sound speed. A similar interaction is also present in the next case with $\beta_3$.
\subsection{Hidden Preheating by $\beta_3(t)$}
\label{sec:beta3}
The quadratic Lagrangian in this case is
\begin{align}L^{(2)}=\int d^3x a^3&\Bigg[\frac{1}{2}\dot{\chi}^2_c+\frac{1}{2}\dot{\pi}_c^2-\frac{(\partial_i\chi_c)^2}{2a^2}-\frac{(\partial_i\pi_c)^2}{2a^2}-\frac{m_\chi^2(t)}{2}\chi_c^2-\frac{m^2_\pi(t)}{2}\pi_c^2\\
&-\frac{\dot{\beta}_3\dot{\chi}_c\pi_c}{\sqrt{-2m^2_{pl}\dot{H}}}-\frac{\ddot{H}}{\dot{H}}\frac{\beta_3\pi_c\dot{\chi}_c}{\sqrt{-2m^2_{pl}\dot{H}}}-\frac{\beta_3\dot{\pi}_c\dot{\chi}_c}{\sqrt{-2m^2_{pl}\dot{H}}}-\frac{\ddot{\beta}_3\pi_c\chi_c}{\sqrt{-2m^2_{pl}\dot{H}}}\Bigg].\end{align}
This time there are three different couplings,
\be \rho_3\equiv\frac{\beta_3}{\sqrt{-2m^2_{pl}\dot{H}}},~~ R_2\equiv\frac{\ddot{\beta}_3}{\sqrt{-2m^2_{pl}\dot{H}}},~~R_3\equiv\frac{\dot{\beta}_3}{\sqrt{-2m^2_{pl}\dot{H}}}.\ee
With these new definitions the quadratic Lagrangian in terms of $\tilde{\chi}_c$ and $\tilde{\pi}_c$ becomes
\begin{align}L^{(2)}=\frac{1}{2}&\int d^3x\Bigg[\dot{\tilde\chi}^2_c+\dot{\tilde\pi}_c^2-\frac{(\partial_i\tilde{\chi}_c)^2}{a^2}-\frac{(\partial_i\tilde{\pi}_c)^2}{a^2}-\tilde{m}_\chi^2(t)\tilde{\chi}_c^2-\tilde{m}^2_\pi(t)\tilde{\pi}_c^2\\
&-2R_3\dot{\tilde\chi}_c\tilde{\pi}_c-2\frac{\ddot{H}}{\dot{H}}\rho_3\tilde{\pi}_c\dot{\tilde\chi}_c-2\rho_3\dot{\tilde\pi}_c\dot{\tilde\chi}_c+3H\rho_3\left(\tilde{\pi}_c\dot{\tilde{\chi}}_c+\tilde{\chi}_c\dot{\tilde{\pi}}_c\right)\\
&3H\left(R_3-\frac{\ddot{H}}{\dot{H}}\rho_3-\frac{9}{2}H^2\rho_3\right)\tilde{\pi}_c\tilde{\chi}_c-2R_2\tilde{\pi}_c\tilde{\chi}_c\Bigg].
\end{align}
The coupling strength $\beta_3$ has mass dimension two. This makes $\rho_3$ dimensionless, just like $\rho_2$. 
$R_2$ and $R_3$ are the dimension full parameters that can set the scales here. The $R_2$ term only contributes to the over all energy. $R_3$ has dimensions of mass and works similar to $R_1$. In the limit $a\to 1$ while the combination $m^2_{pl}\dot{H}$ stays finite, let us assume that $R_2$, $R_3$ and $\rho_3$ stay finite and are constant while the terms involving $3H$ and $\frac{\ddot{H}}{\dot{H}}$ can be dropped. With these assumptions we are focusing on scales $R_3\gg \omega \gg m_\phi >H$ and the Lagrangian equations of motion take the form
\be \ddot{\tilde\chi}_{ck}+\left(k^2+\tilde{m}^2_\chi(t)\right)\tilde{\chi}_{ck}=R_3\dot{\tilde\pi}_{ck}-R_2\tilde{\pi}_{ck}+\rho_3\ddot{\tilde\pi}_{ck},\ee
\be \ddot{\tilde\pi}_{ck}+\left(k^2+\tilde{m}^2_\pi\right)\tilde{\pi}_{ck}=-R_3\dot{\tilde\chi}_{ck}-R_2\tilde{\chi}_{ck}+\rho_3\ddot{\tilde\chi}_{ck}.\ee
This exhibits modes with frequencies of
\begin{align}\nn &\omega_\pm^2=\frac{1}{\left(1-\rho_3^2\right)}k^2+\frac{\tilde{\mathcal{B}}^2}{2\left(1-\rho_3^2\right)}\\
&\pm\frac{1}{2\left(1-\rho_3^2\right)}\sqrt{4\rho_3^2k^4+4\left(\rho^2_3\left(\tilde{m}^2_\chi+\tilde{m}^2_\pi\right)+2\rho_3R_2+R^2_3\right)k^2+\tilde{\mathcal{B}}^4-4\left(1-\rho_3^2\right)\mathcal{M}^4}\end{align}
where
\begin{align} \tilde{\mathcal{B}}^2&\equiv \tilde{m}^2_\chi+\tilde{m}^2_\pi+R^2_3+2\rho_3R_2,\\
\mathcal{M}^4&\equiv \tilde{m}^2_\chi \tilde{m}^2_\pi -R^2_2\end{align}
have been defined for ease of notation.
Inflaton and reheating modes in the range $k^2\gg \tilde{\mathcal{B}}\left( R_3,\tilde{m}_\chi,\tilde{m}_\pi,R_2\right)$, $\mathcal{M}^2\left(\tilde{m}_\chi,\tilde{m}_\pi,R_2\right)$, propagate freely with $\omega\sim c_\rho k$  where they can acquire a sound speed of the order of $c^2_\rho= \frac{1\pm\rho_3^2}{1-\rho_3^2}$. The main difference of this case from the case with $\beta_1$ is the possibility that a sound speed exists even at this relatively high range of energies! This suggests that these type of couplings arise from the presence of at least three fields, where one of the fields is much heavier then both the inflationary and reheating sectors, leading to a nontrivial sound speed, $c_\rho^2$, even at the ranges where the two sectors of interest appear weakly coupled to each other.

In the range $R_3^2\gg \rho_3k^2 > \tilde{m}^2_\chi,\tilde{m}^2_\pi,R_2$, $\tilde{\mathcal{B}}^2\simeq R^2_3$ and the mode frequency is approximately
\be \omega_\pm^2\simeq \frac{k^2}{\left(1-\rho_3^2\right)}+\frac{R_3^2}{2\left(1-\rho_3^2\right)}\pm\frac{R_3^2}{2\left(1-\rho_3^2\right)}\sqrt{1+\left(4\frac{\rho_3^2}{R_3^4}k^4+\frac{4}{R_3^2}k^2\right)}.\ee
Expanding the square root, this range accommodates modes with the dispersion relation
\be \omega^2_+\simeq \frac{R^2_3}{1-\rho^2_3}+\frac{2}{1-\rho^2_3}k^2.\ee
At leading order these are very heavy modes that do not propagate since there is no k-dependence in the first term. The lighter modes in this range have the following dispersion relation
\be \label{omega-_beta3} \omega_-^2\simeq \frac{k^4}{R_3^2}.\ee
Note that this expression is very similar to the dispersion relation \eqref{hiddenbeta1_rkm} that we found for modes in the Hidden preheating regime with $\beta_1$ type couplings, which in that case were purely inflaton modes.


For $\rho_3\ll 1$, in the regime $R_3\gg\omega>m_\phi,R_2>H_m$ the Lagrangian 
\be \label{hiddenlag_beta3} L^{(2)}\simeq \int d^3x\left[ -R_3\dot{\tilde\chi}_c\tilde{\pi}_c-\frac{1}{2a^2}(\partial_i\tilde{\chi}_c)^2-\frac{1}{2a^2}(\partial_i\tilde{\pi}_c)^2-\frac{1}{2}\tilde{m}^2_\chi(t)\tilde{\chi}_c^2-\frac{1}{2}\tilde{m}^2_\pi(t)\tilde{\pi}_c^2\right].\ee
gives rise to the example where the $\tilde{\chi}_c$ modes are the lightest degree of freedom, and $\tilde{\pi}_c$ plays the role of their canonical momenta
\be \label{canmomt_beta3} p_\chi\equiv\frac{\partial\mathcal{L}}{\partial\dot{\tilde\chi}_c}=-R_3\tilde{\pi}_c,~~~~p_\pi\equiv \frac{\partial \mathcal{L}}{\partial\dot{\tilde{\pi}}_c}=0.\ee
And again there is a constraint which now demands that the momentum of inflaton modes vanish.
 
Taking the limit $a\to 1$ with $m_{pl}^2\dot{H}$ remaining finite, and demanding equations \eqref{canmomt_beta3} be satisfied, the Hamiltonian corresponding to \eqref{hiddenlag_beta3} is
\begin{align}
\nn H&=\int d^3x \left[p_\pi\dot{\tilde\pi}_c+p_\chi\dot{\tilde{\chi}}_c\right]-L\\
\label{hamhidden_beta3}&=\frac{1}{2}\int d^3x \left[\left(\partial_i\tilde{\chi}_c\right)^2+\frac{1}{R^2_3}\left(\partial_ip_\chi\right)^2+\tilde{m}_\chi^2\tilde{\chi}_c^2+\frac{\tilde{m}_\pi^2}{R^2_3}p^2_\chi\right].
\end{align}
Decomposing the fields into their Fourier modes, the Hamilton equations of motion read
\begin{align}
\dot{p}_{\pi k}&=-\frac{\partial H}{\partial \tilde{\pi}_{ck}}=0,\\
\dot{\tilde{\pi}}_{ck}&=\frac{\partial H}{\partial p_{\pi k}}=0,\\
\dot{p}_{\chi k}&=-\frac{\partial H}{\partial \tilde{\chi}_{ck}}=-\left[k^2+\tilde{m}_\chi^2\right]\tilde{\chi}_{c k},\\
\dot{\tilde{\chi}}_{ck}&=\frac{\partial H}{\partial p_{\chi k}}= \frac{1}{R^2_3}\left[k^2+\tilde{m}_\pi^2\right]p_{\chi k}.
\end{align}
The first of these guarantee that the constraint is satisfied at all times. With our previous assumptions and neglecting the time dependence in $R_3$, $\tilde{m}_\chi$ and $\tilde{m}_\pi$, this system exhibits modes $\tilde{\chi}_{ck}\sim e^{-i\omega t}$ with the dispersion relation
\be \label{hiddendis_chi_beta3} \omega=\frac{1}{R_3}\sqrt{k^4+\left(\tilde{m}_\chi^2+\tilde{m}_\pi^2\right)k^2+\tilde{m}_\chi^2\tilde{m}_\pi^2}\ee
for scales in the range $R_3\gg k> m_\phi,R_2>H_m$ and $\rho_3\ll1$. In fact at leading order in this regime equation \eqref{omega-_beta3} is recovered
\be \label{hiddenregime_beta3_dis} \omega \simeq\frac{k^2}{R_3}=\omega_-\ee
Thus we can identify $\omega_-$ modes to be purely reheating modes. This is similar to the case of Hidden preheating with $\beta_1$, only this time the roles of the two fields are switched around. Since the reheating perturbations are the light modes here, these type of kinetic couplings could be more likely to give rise to efficient preheating. Unlike the case with $\beta_1(t)$, there do not seem to appear previously studied examples to this case in neither the inflationary nor preheating literature. This may be due to the fact that in generalizing couplings usually the main attention is given to modifying the kinetic terms of the inflaton.    

The scale that the $R_3$ defines, the scale up to which $\chi_c$ is the single effective species, is around $\mathcal{O}\left(\frac{m_{pl}H}{m_\phi}\right)$ order of magnitude wise. Considering $\sqrt{m_{pl}H}$ as a unit scale, the relationship between the magnitude of this scale to the $\chi$-production scale is $K_{bck}=(m_{pl}H)^{1/3}R_3^{1/3}\sim R_3^{1/3}$. 
In conclusion at frequencies below the $R_3$ scale, $\omega<R_3$, the effective modes are the reheating modes alone where as at scales $R_3<\omega<K_{bck}$ modes of both $\pi_c$ and $\chi_c$ are present, and above $K_{bck}$ there is only the inflaton perturbations due to the lack of resonant $\chi_c$ production. These scales and the corresponding species are summarized in figure 2.
\begin{figure}[h]
	\label{beta3modes}
	\includegraphics[width=\textwidth]{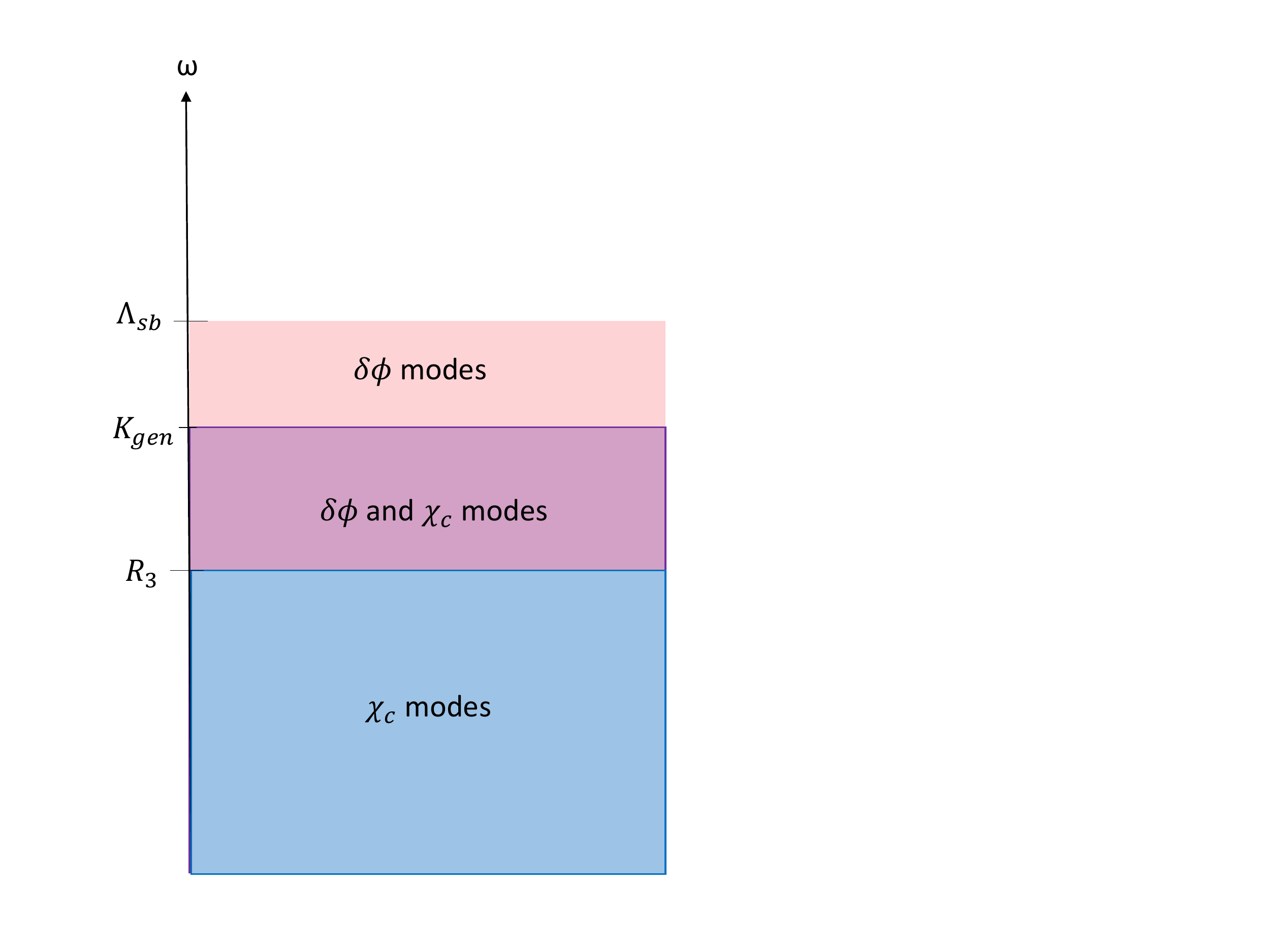}
	\caption{Here the regions where the inflaton modes $\delta\phi=\pi_c$ and reheating modes $\chi_c$ appear as effective degrees of freedom in the presence of $\beta_3$ interactions, where a sound speed is probable at all levels, are shown.}
\end{figure}

\section{Conclusions}
\label{sec:conc}

Low energy effective field theories (EFT), especially the ones that are developed at the level of perturbations such as the EFT considered here, aim to capture the variety of interactions in the most general way. This generality is achieved by considering all of the interactions allowed by the symmetries that are present below a specified scale. This scale in the EFT set up considered here was the scale of spontaneous breaking of time translation invariance, due to the time dependent nature of the background $H(t)$. Among the possible interactions for the inflaton and reheating field perturbations, the present work has focused on the extra derivative couplings that appear under three different classes, specified by the EFT parameters $\{\beta_1,\beta_2,\beta_3\}$. The scales these derivative couplings introduce, the nature of the effective degrees of freedom at energies below the introduced scales ( whether they are inflaton perturbations or the scalar reheating perturbations), and the corrections to the dispersion relation for the effective modes at low energies have been explored in this work. The properties of the background as determined by the preheating era, that is the presence of two scales $m_\phi$ and $H_m$ with the hierarchy $m_\phi\gg H_m\simeq H_p$ between them, have been used to determine the hierarchy between the scales of the interactions, such as the particle production scale and the scale associated to derivative couplings. All this has led to the main conclusion that, all though the aim of preheating is to capture energy transfer between two different species, here the inflaton and the scalar reheating sector $\chi$, in the presence of such derivative couplings only one of the species propagates as an effective degree of freedom at very low energies, while the other stays hidden and modifies the dispersion relation of the propagating species. Instead of an analysis of instability bands to determine the efficiency of $\chi$-production, the main pursuit here has been the identification of the relevant species for low energies and exploring how the dispersion relation of this species gets modified. It is left for future to discuss the efficiency in production of the identified low energy modes through studying the details of resonance in comparison to perturbative decay rates.

While this EFT method allows one to study the properties of perturbations right away, the disadvantage can be that it is not always clear what kind of interactions at the background level would give rise to these interactions at the level of perturbations. For example, an interaction of type $(\frac{1}{\Lambda}\partial^\mu\phi\partial_\mu\phi)X$ where $\phi$ is the inflaton which is to be expanded as $\phi(\vec{x},t)=\phi_0(t)+\delta\phi(\vec{x},t)$, $X$, with $X(\vec{x},t)=X_0(t)+\chi(\vec{x},t)$, is the reheating field and $\Lambda$ is some mass scale; is an example that gives rise to $\beta_1$ type couplings. And indeed these type of interactions are common in inflationary literature in many studies that wish to respect the shift symmetry for the inflaton. On the other hand, $\frac{1}{\Lambda}(\partial^\mu X\partial_\mu X)\phi$ would be an example to $\beta_3$ type couplings, which however would not be an interaction to consider if one is concerned with a shift symmetric inflaton. Looking at the preheating literature, the interactions considered are more of a polynomial type, for instance $g^2\phi^2X^2$ is the first case that has been considered. Derivative couplings during preheating have not been studied at the level they have been during inflation. So far, derivative couplings in preheating literature involve examples of only the class of $\beta_1$ couplings, among the three classes that the EFT methods suggest. Moreover the examples to $\beta_1$ case that have been studied are noted to be not very efficient for resonant production of low energy reheating modes, $\tilde{\chi}_{ck}$. Looking at the dispersion relations, here it is noted that at scales below the scale of derivative coupling $R_1$, the reheating modes appear to effect the canonical momenta of the inflaton perturbations, leaving them as the low energy species with a modified dispersion relation derived in equations \eqref{hiddenbeta1_rkm} and \eqref{omega-smallest}. On the other hand, $\beta_3$ interactions accommodate the reheating modes as the light degrees of freedom with a modified dispersion relation of \eqref{hiddenregime_beta3_dis}. Hence these later type of interactions may be more promising for resonant production of $\chi$ through derivative couplings. 
 
 Moreover some of the derivative couplings, in the presence of $\beta_2$ and $\beta_3$ imply a sound speed and modified dispersion relations for both of the species even at energies where modes of both propagate freely. This suggests that these EFT coefficients may address models that involve additional heavy degrees of freedom. 

The reheating sector $\chi$ as considered here is quite general. Being a primordial scalar field, $\chi$ is most likely to contribute to structure formation and resemble fields associated to dark matter. In principal any of these couplings can arise in models of multi-field inflation. Since the effective field theory method at the level of perturbations followed here considers all possible terms that respect the symmetries at the scales of interest, one of the expected benefits of this approach is to come across new types of interactions that may not have been thought of yet. The appearance of the less explored case of $\beta_3(t)$ type couplings are an example to this. They would arise from attempts of generalizing interactions of the reheating field, as sketched in the previous paragraphs. It is left for future to explore for this later case, the phenomenological implications and the detailed structure of resonance in comparison with rate of perturbative decays.
With regards to perturbative decay rates, the EFT interactions would give the possible Feynmann diagrams to be computed, however the strength of the amplitude from these diagrams will depend on the coupling parameter which in turn depends on the background physics. The background information is determined by how $H_p(t)$ and $m_\phi$ work into $\phi_0(t)$ and $X_0(t)$. The same holds for the efficiency of resonant particle production. One can make estimates on the scale of particle production from the general behavior of the background as it has been done here, but to study the actual efficiency one again has to first study the details of how the background parameters work into the Mathieu variables. Once solid examples that give rise to $\beta_3$ type interactions at the level of perturbations are constructed, then how the coupling parameters depend on $m_{pl},$ $H$ and $m_\phi$ through the background behavior of $\phi_0$ and $X_0$ will become more clear and, the perturbative decay rates and efficiency of resonance can be studied more concretely. 
\begin{acknowledgements}
	It is a great pleasure to thank Scott Watson and an anonymous referee for their valuable comments on earlier versions of this manuscript, to Cristian Armendariz Picon, Johanna Karouby, Caner Ünal, and Evangelos Sfakianakis for useful discussions, to Vak\i f \"Onemli and Costantinos Skordis for their encouragement. This work was initiated at Syracuse University Physics Department under the Graduate Student Assistantship, supported in part by the DOE grant DE-FG02-85ER40237 and completed at CEICO supported by the IOP Researchers Mobility Grant $CZ.02.2.69/0.0/0.0/16\_027/0008215$.	
	
\end{acknowledgements}
\appendix
\section{Behavior of the Background in the Early Stages of Preheating}
\label{ap:background}
As noted in the introduction, our starting point is that at the background level the energy momentum density is dominated by a single scalar field, $\phi_0$. This scalar field is same as the one that dominated the energy momentum density during inflation, the inflaton, and it only exhibits time dependence $\phi_0(t)$. During inflation, its time dependence is weak. If this background scalar is minimally coupled to gravity,
\be S=\int d^4x\sqrt{-g}\left[\frac{1}{2}m^2_{pl}R-\frac{1}{2}g^{\mu\nu}\partial_\mu\phi_0\partial_\nu\phi_0-V(\phi_0)\right],\ee
what will be the behavior of the overall background $H_p(t)$, at the end of inflation when the slow roll conditions no longer hold? 

Assuming that $\phi_0(t)$ minimizes its potential, the leading term in its potential will be the mass term $V(\phi_0)\sim m^2_\phi\phi_0^2$. On an FLRW background, the scalar field evolves according to
\begin{align}
\ddot{\phi}_0+3H\dot{\phi}_0+V'(\phi_0)=0.
\end{align}
The time derivative of the scalar field can be considered to be $\dot{\phi}_0\sim m_\phi\phi_0$. If the friction term $H\dot{\phi}_0$ is neglected, the scalar field evolution would be $\phi_0(t)\sim\Phi sin(m_\phi t)$. The effect of the Hubble friction gives further time dependence to the amplitude $\Phi$. 

The evolution of the Hubble parameter is governed by
\begin{subequations}
	\begin{align} \label{h2}6m^2_{Pl}H^2=\dot{\phi}_0^2+m_\phi^2\phi_0^2,\\
	\label{hdot}2m^2_{Pl}\dot{H}=-\dot{\phi}^2_0.
	\end{align}
\end{subequations}
Following \cite{Mukhanov:2005sc}, let us switch from the variable $\phi_0$ to $\theta$ defined as
\begin{subequations}
	\begin{align}
	\label{theta1}\phi_0=\sqrt{6}m_{Pl}\frac{H}{m_\phi}sin\theta,\\
	\label{theta2}\dot{\phi}_0=\sqrt{6}m_{Pl}Hcos\theta.
	\end{align}
\end{subequations}
This definition automatically satisfies \eqref{h2} and gives
\be \dot{H}=-3H^2cos^2\theta.\ee 
As an internal consistency the derivative of \eqref{theta1} should give \eqref{theta2}. This condition leads to
\be \dot{\theta}=m_\phi+\frac{3}{2}Hsin(2\theta).\ee
For the era under consideration $H_p\ll m_\phi$, and hence $\theta\simeq m_\phi t+\Delta$. Using this approximation in \eqref{hdot} gives 
\be -\int^{H(t)}_{H_{end}}\frac{dH}{H^2}=3\int^{t}_{t_{end}}cos^2(m_\phi t'+\Delta)dt',\ee
which is to be solved for H. The end of inflation occurs when 
\be \epsilon(t_{end})=\frac{-\dot{H}_{end}}{H^2_{end}}=1.\ee 
From equations \eqref{h2} and \eqref{hdot} with $\phi_0(t_{end})\sim m_{Pl}$, this means $H_{end}\simeq \frac{m_\phi}{2}$. The solution for H(t) reads as
\begin{align} \nn H_p&=\frac{2}{3t}\left[1+\frac{sin(2m_\phi t+2\Delta)}{2m_\phi t}\right]^{-1}\\
&\label{Hp}\simeq H_m\left[1-\frac{3H_m}{4m}sin(2m_\phi t+2\Delta)+\frac{9}{16}\left(\frac{H_m}{m_\phi}\right)^2sin^2(2m_\phi t)+....\right]\end{align}
where $\alpha \equiv\frac{sin(2m_\phi t+2\Delta)}{2m_\phi t}$ is small at times $1\ll m_\phi t$ and hence one can consider a series expansion around $\alpha=0$. So the end of inflation represents a matter dominated era with oscillatory corrections.\footnote{The presence of these oscillatory corrections is what makes an era dominated by oscillating scalar field different then an era of dust which behaves exactly as $H_{dust}=H_m$ with zero pressure.} Equation \eqref{theta1} then, gives the following behavior for the inflaton
\be\label{phibackg}
\phi_0(t)\simeq \sqrt{6}m_{pl}\frac{H_m}{m_\phi}\left[sin(m_\phi t)+\frac{3}{8}\frac{H_m}{m_\phi}(cos3m_\phi t-cosm_\phi t)+...\right].\ee
The behavior of the inflaton perturbations on this background and the duration of this oscillatory era have been studied to understand the end of single field inflation with canonical kinetic term and minimal coupling to gravity \cite{Easther:2010mr,Jedamzik:2010dq}. The original example of preheating \cite{Kofman:1997yn} considers only the zeroth order terms in this background 
\begin{subequations}
	\begin{align}
	H_{pc}&=H_m,\\
	\phi^{pc}_0(t)&=\sqrt{6}m_{Pl}\frac{H_m}{m_\phi}sin(m_\phi t+\Delta)
	\end{align} 
\end{subequations}
and involves a second field $\chi$ to which the inflaton transfers its energy via the coupling $g^2\phi^2\chi^2$. 

\bibliography{references}
\bibliographystyle{plain}

\end{document}